# A bright triggered twin-photon source in the solid state

T. Heindel[1,†,*], A. Thoma[1,†], M. von Helversen[1], M. Schmidt[1,2], A. Schlehahn[1], M. Gschrey[1], P. Schnauber[1], J.-H. Schulze[1], A. Strittmatter[1,3], J. Beyer[2], S. Rodt[1], A. Carmele[4], A. Knorr[4], and S. Reitzenstein[1]

[1]Institut für Festkörperphysik, Technische Universität Berlin, 10623 Berlin, Germany
[2]Physikalisch-Technische Bundesanstalt, Abbestraße 1, 10587 Berlin, Germany
[3]Present Adresse: Institut für Experimentelle Physik, Otto-von-Guericke Universität Magdeburg, PF4120, Magdeburg, Germany
[4]Institut für Theoretische Physik, Technische Universität Berlin, 10623 Berlin, Germany
†These authors contributed equally to this work.
*e-mail: tobias.heindel@tu-berlin.de

**Abstract** A non-classical light source emitting pairs of identical photons represents a versatile resource of interdisciplinary importance with applications in quantum optics and quantum biology. Emerging research fields, which benefit from such type of quantum light source, include quantum-optical spectroscopy or experiments on photoreceptor cells sensitive to photon statistics. To date, photon twins have mostly been generated using parametric downconversion sources, relying on Poissonian number distributions, or atoms, exhibiting low emission rates. Here, we propose and experimentally demonstrate the efficient, triggered generation of photon twins using the energy-degenerate biexciton-exciton radiative cascade of a single semiconductor quantum dot. Deterministically integrated within a microlens, this nanostructure emits highly-correlated photon pairs, degenerate in energy and polarization, at a rate of up to $(234 \pm 4)$ kHz. Furthermore, we verify a significant degree of photon-indistinguishability and directly observe twin-photon emission by employing photon-number-resolving detectors, which enables the reconstruction of the emitted photon number distribution.

To realize an integrated light source capable of emitting non-classical multi-photon states, is a fascinating, yet equally challenging task at the heart of quantum optics[1-4]. Emerging research fields, which benefit from such type of quantum light source, include quantum-optical spectroscopy[5-7] or experiments on photoreceptor cells sensitive to photon statistics[8]. To date, photon twins have mostly been generated using parametric downconversion sources[9], relying on Poissonian number distributions, or atoms[10,11], exhibiting low emission rates. Integrated schemes using parametric downconversion for the generation of photon twins have been demonstrated[12], but still suffer from low efficiencies. Semiconductor quantum dots (QDs), on the other hand, turned out to be excellent quantum emitters[13-15], which can produce close to ideal single-photon states with high efficiency under optical[16-18] and electrical[19] excitation. Typical photon-pair experiments exploiting the biexciton-exciton radiative cascade in QDs[20] aim at the generation of entanglement[21,22]. These experiments entirely relied on pairs of photons with different energies, which eliminated the possibility to directly generate photon-twins.

Here, we propose and experimentally demonstrate an integrated source of photon twins, e.g. pairs of photons with same energy and polarization highly correlated in time. For this purpose, we use a QD exhibiting an energy degenerate biexciton-exciton radiative cascade integrated deterministically within a monolithic microlens employing 3D in-situ electron-beam lithography. Twin-photon emission of our quantum light source is first studied and verified via polarization-resolved photon-correlation measurements. Additionally, we verify a significant degree of photon-indistinguishability in Hong-Ou-Mandel-type two-photon interference experiments. To directly observe the twin-photon emission of our source, we further employ a photon-number-resolving detector based on a transition edge sensor, which enables us to reconstruct the photon number distribution emitted by the twin-photon source and to compare the result with a QD-based single-photons source.

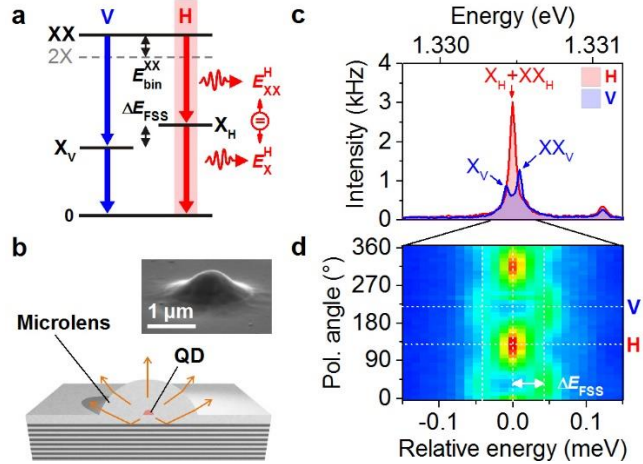

**Figure 1 | Concept of a deterministically integrated twin-photon source.** (**a**) Energy level scheme of a radiative cascade involving the biexciton- (XX), exciton- (X) and ground- (0) state. For finite fine structure splitting $\Delta E_{FSS}$, the possible decay channels are linear-horizontally (H) and -vertically (V) polarized. In case of $E_X^H = E_{XX}^H$ the exciton fine structure splitting $\Delta E_{FSS}$ equals the biexciton binding energy $E_{bin}^{XX}$ and the photons within the H-polarized decay channel exhibit identical energy and polarization. (**b**) Illustration of our solid-state based quantum light source constituted of a single quantum dot (QD) deterministically integrated within a monolithic microlens. The microlens design in combination with a lower distributed-Bragg reflector allows for an enhanced photon collection efficiency of photons emitted by the QD. Inset: Scanning electron microscope image of a microlens. (**c**) Spectrally resolved photoluminescence of a single-QD microlens for H- and V-polarization. For H-polarization the superimposed emission of exciton and biexciton leads to an increased emission intensity compared to V-polarization. (**d**) Polarization-resolved emission spectra in a close up with relative energy scale. A quantitative analysis reveals $\Delta E_{FSS} = |E_{bin}^{XX}| = (51 \pm 6)$ µeV. By selecting the H-polarized decay channel photon-twins can be extracted.

## Results

**Concept of the deterministic twin-photon source.** The biexciton state of a QD is constituted of two electron-hole pairs and typically, due to Coulomb and exchange interactions of the involved charge-carriers, shows a finite binding energy $E_{bin}^{XX}$ with respect to the case of the two unbound excitons ($E_{bin}^{XX} \sim 1$ meV in case of the InGaAs/GaAs material system[23]). The exciton state, on the other hand, consists of a single electron-hole pair, and usually reveals a fine structure splitting $\Delta E_{FSS}$ (~10 µeV[24]) small compared to $E_{bin}^{XX}$, which arises from anisotropic electron-hole exchange interactions. The resulting radiative cascade emits pairs of photons in two possible decay channels, one being linear-horizontally (H) and the other one linear-vertically (V) polarized. Due to the energy scales of $E_{bin}^{XX}$ and $\Delta E_{FSS}$ mentioned above, this configuration leads to two doublets of orthogonally linearly polarized emission lines visible in the emission spectra of exciton and biexciton state exhibiting spectrally distinguishable photons. In this work, we selected a QD featuring $E_X^H = E_{XX}^H$ (cf. Fig. 1a), which is a direct consequence of $\Delta E_{FSS} = |E_{bin}^{XX}|$ for the chosen radiative cascade. For this particular energy level alignment, one decay channel of the XX-X cascade reveals the emission of photon-twins – a light state constituted of two temporally correlated photons with identical emission energy and polarization. Additionally, we utilized 3D in-situ electron-beam lithography[25], to integrate the QD deterministically within a monolithic microlens (cf. Fig. 1b), which provides enhanced photon collection efficiency for the twin-photon generation process (see Methods). Figure 1c shows photoluminescence spectra of the QD's emission under above-bandgap ($\lambda = 850$ nm) continuous wave excitation for linear-horizontal (H) and -vertical (V) polarization. In case of V-polarization a doublet centered at 1.33047 eV is observed, where the low- and high-energy component can be attributed to the excitonic ($X_V$) and biexcitonic ($XX_V$) emission, respectively (cf. Fig. 1a). Switching to H-polarization, a single, intense emission line can be observed at 1.33047 eV. This behavior is analyzed in more detail in Fig. 1d, depicting a polarization-resolved map of photoluminescence spectra. Exciton and biexciton exhibit a sinusoidal shift in energy with opposite phase, however, close to H polarization their emission becomes superimposed, resulting in a distinct maximum of the emission intensity. A quantitative analysis of the spectra from Fig. 1d yields $\Delta E_{FSS} = |E_{bin}^{XX}| = (51 \pm 6)$ μeV (see Supplementary Section 1, Fig. S1). Here, the fact that we observe an antibinding biexciton state ($E_{XX}^V > E_X^V$), which occurs in ~5% of the QDs in this sample, is indicative for a relatively small QD size[26].

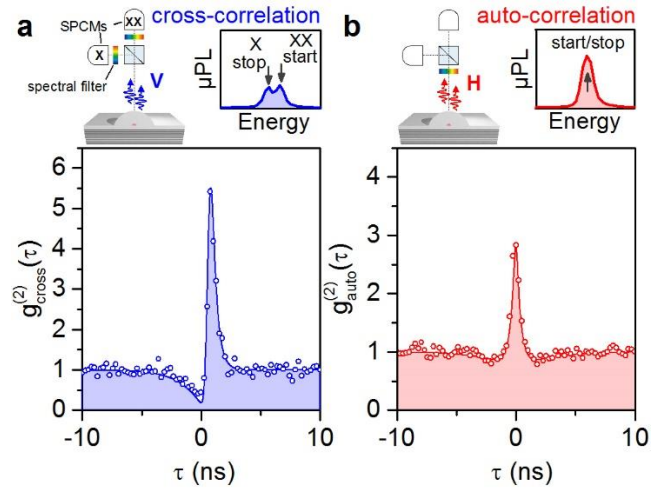

**Figure 2 | Polarization-resolved photon-correlations of photon pairs emitted by the quantum dot microlens under continuous wave excitation.** (a) Photon cross-correlation histogram $g^{(2)}_{cross}(\tau)$ for the V-polarized decay channel, where biexciton and exciton emission are spectrally separated (cf. schematic of experiment and spectrum). The strong bunching signature at $\tau > 0$ in combination with an antibunching at $\tau < 0$ proves the cascaded emission of XX-X photon pairs. (b) Photon auto-correlation histogram $g^{(2)}_{auto}(\tau)$ for the H-polarized decay channel of the XX-X cascade, where biexciton and exciton emission is superimposed (cf. schematic). The pronounced bunching at $\tau = 0$ of $g^{(2)}_{auto}(0) = 2.85$ indicates a high degree of two-photon correlations, due to the emission of photon twins. Solid curves in both panels are theoretical simulations based on a four-level master equation approach accounting for the experimental conditions.

**Polarization-resolved photon-correlations.** The dynamics of this unique four-level system were studied via polarization-resolved photon-correlation measurements[27]. First, we address the correlations of the V-polarized cascade channel. In this case, X and XX photons are energetically separable using two spectrometers (cf. schematic in Fig. 2a). Figure 2a displays the obtained cross-correlation coincidence histogram $g^{(2)}_{cross}(\tau)$, where XX-photons started and X-photons stopped the measurement. An asymmetric bunching effect is observed for positive delay times $\tau$, owing to the cascaded emission of photon-pairs within the same decay channel[28,29]. Next, the photon correlations within the H-polarized decay channel are investigated. Here, exciton and biexciton photons are energetically degenerate and temporal correlations can be probed via photon auto-correlation measurements using a single spectrometer (cf. schematic in Fig. 2b). The corresponding coincidence histogram reveals a prominent bunching signature at zero delay and - due to the absence of time-ordering of the detected photons - a symmetric behavior in $\tau$. The pronounced bunching indicates a high degree of two-photon correlations, which proves that this unique XX-X radiative cascade serves as a source of photon twins. Additionally, we detect clear antibunching at finite delay times ($\tau = \pm 2$ ns), signifying the non-classicality of the emitted light state. Our experimental observations agree quantitatively with a theoretical model (solid curves) based on a four-level rate equation approach (see Supplementary Section 2).

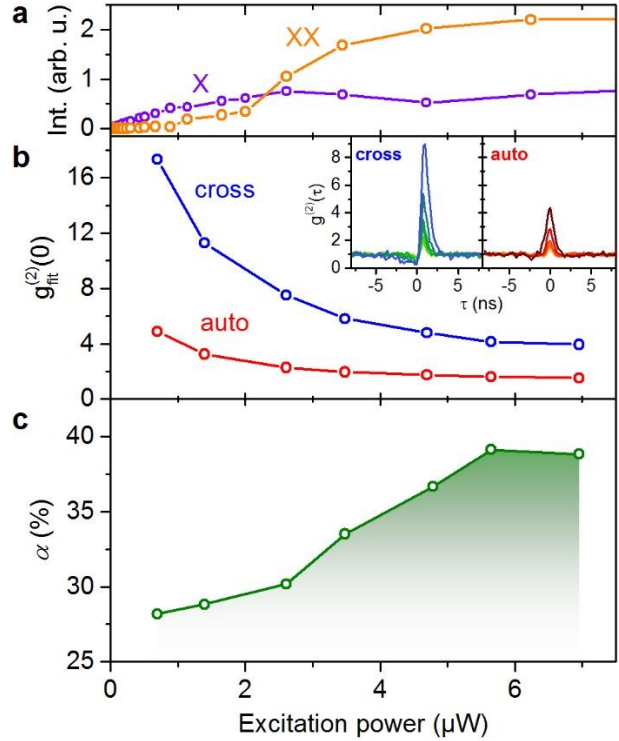

**Figure 3 | Excitation power dependence of twin-photon generation.** (**a**) Integrated intensities of exciton and biexciton emission extracted for the V-polarized decay channel. The emission intensities of X and XX saturate at excitation powers of about 2 μW and 6 μW, respectively. (**b**) Bunching values $g^{(2)}_{fit}(0)$ for auto- and cross-correlation resulting from a theoretical fit to the experimental data shown in the inset (taking into account the setup's timing resolution). (**c**) Fraction of two-photon correlations due to twin-photon emission $\alpha = g^{(2)}_{auto}(0)/g^{(2)}_{cross}(0)$ calculated from the bunching values in (**b**).

**Quantifying the source efficiency.** The magnitude of the bunching in Fig. 2 itself, however, does not carry information about the probability of having a photon pair per excitation. In fact the bunching value depends mainly on the occupation of the exciton level (inversely proportional), rather than on the biexciton occupation. For this reason and to quantify the efficiency of our source, we introduce the parameter $\alpha = g^{(2)}_{auto}(0)/g^{(2)}_{cross}(0)$, i.e. the ratio of the bunching values in auto- and cross-correlation, respectively. The parameter $\alpha$ thereby corresponds to the fraction of two-photon correlations due to twin-photon emission, which naturally follows if one considers that the observable of our photon auto-correlation measurement on the H-polarized decay channel results from the superposition of a total of four different photon correlations (see Methods for details and Supplementary Section 2 for the explicit expressions of the observables). Hence, by comparing the measured cross- and auto-correlation traces, one obtains information about the efficiency of the twin-photon cascade. Figure 3a displays the excitation power dependencies of the integrated intensities of XX and X emission (extracted in V-polarization). The corresponding bunching values $g^{(2)}_{fit}(0)$ are depicted in Fig. 3b and result from a deconvolution of the measured auto- and cross-correlation traces (cf. Fig. 3b, inset), by applying our theoretical model and taking into account the setup's timing resolution. The extracted bunching magnitude reveals a monotonic drop with increasing excitation. This behavior is typically observed in excitation-power dependent cross-correlation measurements[30,31], and does not carry information on the twin-photon generation efficiency (as discussed above). However, the decrease of the bunching for photon-twins (auto-correlation) is less pronounced compared to the distinguishable XX-X photon-pairs (cross-correlation), which indicates a change in the generation efficiency of photon twins in the degenerate cascade channel. Figure 3c presents the respective ratio α calculated from Fig. 3b. With increasing excitation, cascade efficiency steadily increases and reaches a maximum value of α = (39 ± 3)%. From this, we can deduce the twin-photon emission rate *TPR* collected via the microscope objective to be (234 ± 4) kHz (see Methods). This represents a significant improvement (x5) compared to photon twins generated with atoms[11]. As the outcoupling of a photon twin depends quadratically on the photon-extraction efficiency of the microlens, we anticipate further improved *TPRs* of ~1.3–2.1 MHz using anti-reflection coatings[32] or a bottom gold mirror[25] (assuming photon extraction efficiencies of 50–80% and excitation at λ = 850 nm).

**Triggered generation of photon twins and photon indistinguishability.** To operate our quantum light source as two-photon gun, we applied pulsed excitation in the following. Figure 4a displays the auto-correlation histogram of the energy-degenerate decay channel (H-polarization) under above-band pulsed excitation (λ = 850 nm, 80 MHz). Here, the strong bunching effect proves the predominance of two-photon correlations due to pulsed twin-photon emission. Next, pulsed twin-photon emission was utilized to test the indistinguishability of both photons emitted within the H-polarized decay channel, by means of Hong-Ou-Mandel (HOM) -type two-photon interference (TPI) experiments (see Methods). For this purpose, we excited the QD into its *p*-shell (λ = 904.5 nm) and sent the triggered photon-twins into a symmetric Mach-Zehnder interferometer, where a λ/2-waveplate within one interferometer arm allows for switching between co- and cross-polarized measurement configuration (cf. Fig. 4b, illustration). Figure 4b presents the TPI histograms $g^{(2)}_{HOM}(\tau)$ for both measurement configurations. A clear reduction in coincidences is observed at τ = 0 for co-polarized measurement configuration, proving the interference of photons emitted by the cascade. Considering the bunching values $g^{(2)}_{\parallel}(0)$ and $g^{(2)}_{\perp}(0)$ extracted from the coincidence peak-area ratios yields a visibility of two-photon interference of $V = \frac{g^{(2)}_{\perp}(0) - g^{(2)}_{\parallel}(0)}{g^{(2)}_{\perp}(0) \times 1/2}$ =(56 ± 9)% (see Methods). Limiting factors for the observed indistinguishability are discussed in the Supplementary Section 3 and can be encountered by applying strictly resonant excitation schemes via simultaneous one- and two-photon excitation or exploiting cavity effects.

To directly detect the generated twin-photon state, we employed a state-of-the-art photon-number-resolving (PNR) detection system based on a transition-edge sensor (TES) (see Methods). Figure 4c presents the measured photon number distribution of the H-polarized decay channel, triggered by a pulsed diode laser at a repetition rate of 1 MHz (λ = 661 nm). Detection events corresponding to photon-twins ('2') are clearly identified and well separated from the single-photon detection events. Furthermore, we can derive the photon number distribution emitted from our source, by taking into account the total losses of the experimental setup (see Methods). As illustrated in Fig. 4d, the probability for twin-photon emission is $p_{twin} = (8.0 ^{+3.1}_{-1.8})$% in case of the degenerate cascade channel, being noticeably larger than the probability for single-photon emission $(6.2 ^{+1.2}_{-2.0})$%. As expected, these probabilities completely change in case of a single-photon source, showing $(22.7 ^{+5.4}_{-3.8})$% single-photon emission and only $(2.6 ^{+2.3}_{-0.9})$% photon pair emission. In both cases, the relatively large contribution of the vacuum state is partly related to the less efficient excitation using the pulsed diode laser (661 nm) utilized for the PNR

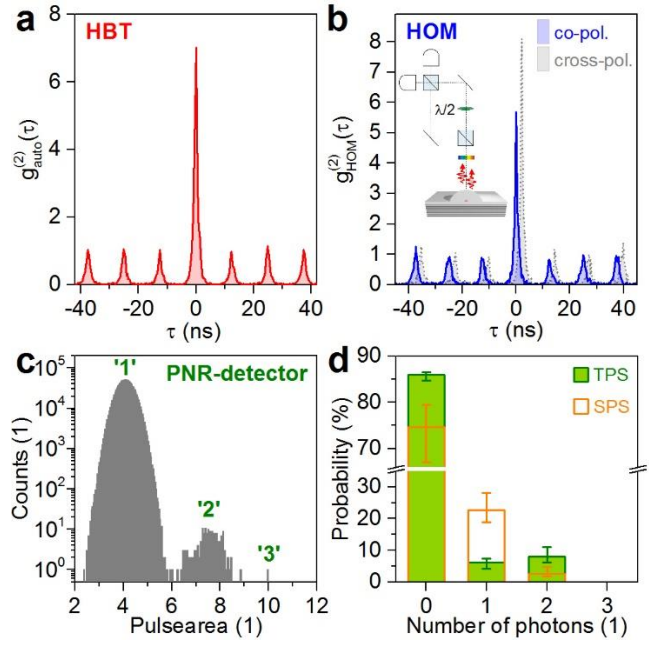

**Figure 4 | Triggered generation of photon twins (H-polarization).** (**a**) Measured photon auto-correlation histogram $g^{(2)}_{auto}(0)$ of photon-twins under pulsed above-band excitation at low excitation power ($P = 87$ nW). A bunching of $g^{(2)}_{auto}(0) = A_0/A = 5.1$ is observed, where $A_0$ and $A$ correspond to the zero-delay peak and the peaks at finite $\tau$. (**b**) Hong-Ou-Mandel (HOM) two-photon interference experiment using photon-twins under pulsed $p$-shell excitation. Measured histograms of $g^{(2)}_{HOM}(\tau)$ for co-polarized (solid line) and cross-polarized (dashed line) configuration using a symmetric Mach-Zehnder interferometer (data in cross-polarization shifted by +2 ns for clarity). (**c**) Direct detection of photon-twins using a photon-number-resolving (PNR) detection system based on a transition-edge sensor (TES). The histogram shows the pulsearea-distribution of photon detection events from the twin-photon cascade, where the labeled peaks correspond to the detection of one ('1'), two ('2') or three ('3') photons. (**d**) Reconstructed photon number distribution of the twin-photon source (TPS), deduced from the PNR measurement in (**c**) by taking into account the total losses of the experimental setup. Results obtained for a single-photon source (SPS) are displayed for comparison.

PNR experiment. In case of the twin-photon source, additionally the population of the V-polarized decay channel is not detected due to the selection of the H-polarization, which artificially increases the contribution of the vacuum state. From the twin-photon emission probability $p_{twin}$, we can further deduce the triggered twin-photon emission rate *TPR* collected via the microscope objective to be ($52^{+21}_{-12}$) kHz (see Methods). Using a more efficient excitation scheme at a wavelength of 850 nm, would thereby readily allow for a threefold enhancement of the *TPR*.

## Discussion

The probability to observe the energy level alignment presented in our work for an as-grown QD is below 1%. Importantly, the device yield for twin-photon sources can be significantly improved in future by applying an advanced in-situ fabrication technique involving a detailed precharacterization[33], where QDs with particular small biexciton binding energy can be pre-selected. Beyond that, even a fully scalable device concept is within reach employing existing technologies. For instance, employing external strain-tuning via piezo-actutators[34], the biexciton binding energy can be precisely adjusted, which would substantially increases the device yield. Furthermore, our scheme can also be extended towards electrical control or current injection via electrical gates[35].

Additionally, resonant excitation, which is typically applied for the coherent excitation of single excitonic states, would be particularly interesting in the presented case of a degenerate XX-X cascade, where the laser coherently drives two quantum emitters: The biexciton state via two-photon excitation[36] (TPE) and the exciton state via one-photon excitation, at which the $X_H$-level is congruent with the virtual intermediate level of the TPE process. In such case, it could be possible to achieve a biexciton occupation close to unity, which would greatly enhance the parameter $\alpha$, i.e. the fraction of two-photon correlations due to twin-photon emission, and hence boost the efficiency of the twin-photon source. As the TPE depends quadratically on the excitation power, this can be tested in the regime of strong pumping (requiring a multiple-$\pi$-Pulse), at which TPE processes dominate the single-photon excitation of the exciton level. It might be even possible in this regime, to produce coherently excited two-photon states which do not reveal the cascade's intrinsic time-ordering, as recently observed in experiments on the dressing of the biexciton state in QDs[37]. This could finally lead to the realization of efficient sources of close-to-ideal two-photon Fock-states $|2\rangle$ on a fully scalable technology platform.

In summary, we introduced a novel type of integrated twin-photon source based on a QD deterministically integrated within a monolithic microlens. Triggered generation of photon pairs with same energy and polarization becomes possible by utilizing a biexciton-exciton radiative cascade, where the biexciton binding energy equals the bright exciton's fine structure splitting. The proposed quantum light source is very attractive for novel quantum optics experiments, such as the excitation of quantum-degenerate quasi-particle states[5].

## Acknowledgments

We acknowledge support from the German Research Foundation (DFG) via the SFB 787 "Semiconductor Nanophotonics: Materials, Models, Devices" and Grant RE2974/9-1, the German Federal Ministry of Education and Research (BMBF) via the VIP-project QSOURCE (Grant No. 03V0630). Parts of the results in this paper come from the project EMPIR 14IND05 MIQC[2]. This project has received funding from the EMPIR programme co-financed by the Participating States and from the European Union's Horizon 2020 research and innovation programme. A. C. gratefully acknowledges support from the SFB 910: "Control of self-organizing nonlinear systems". We thank A.E. Lita and S.W. Nam for providing the TES detector chips and R. Schmidt, E. Schlottmann, F. Gericke and M. Schlösinger for technical assistance.

## Author contributions

T.H., A.T., and A. Schlehahn performed the spectroscopy and correlation experiments and analyzed the experimental data. A.T., M.H. and M.S. performed the photon-number-resolving experiments under supervision of J.B.. M.G. and P.S. performed the CL lithography under supervision of S. Rodt and processed the samples. J.-H.S. and A. Strittmatter grew the samples. A.C. and A.K. performed the theoretical modeling. T.H. and A.T. wrote the manuscript with input from all authors. T.H. conceived the experiment and supervised the project together with S. Reitzenstein. All authors participated in scientific discussions. T.H. and A.T. contributed equally to this work.

## Methods

**Sample.** The QD sample utilized for our experiments was grown by metal-organic chemical vapor deposition (MOCVD) on GaAs (001) substrate. Self-organized InGaAs QDs are deposited above a lower distributed Bragg reflector (DBR) constituted of 23 alternating $\lambda/4$-thick bi-layers of AlGaAs/GaAs. On top of the QD layer, a 400 nm-thick GaAs capping layer provides the material for the subsequent microlens fabrication. Deterministic single-QD microlenses were processed via 3D in-situ electron-beam lithography based on cathodoluminescence (CL) spectroscopy[25]. Shallow hemispheric-section-type microlenses with heights of 400 nm and base widths of 2.4 μm have been chosen, allowing for a photon extraction efficiency of up to 29%[38].

**Experimental setup.** For the micro-photoluminescence (μPL) investigations, the sample is mounted onto the cold-finger of a liquid-Helium-flow cryostat and held at a temperature of $T = 6$ K. The QD microlens is optically excited using a tunable Ti:sapphire laser operating in continuous wave (CW) or pulsed picosecond mode ($f = 80$ MHz). Photoluminescence is collected via a microscope objective with a numerical aperture (NA) of 0.4 serving as first lens of the detection system. The μPL signal is spectrally analyzed using a grating spectrometer with an attached charge-coupled device camera enabling a spectral resolution of 25 μeV. Two-photon emission of the deterministic QD microlens is further studied via polarization-resolved photon-correlation experiments. In case of photon auto-correlation measurements, the superimposed exciton and biexciton emission (H-polarization) is spectrally selected using a single monochromator and analyzed using a fiber-based Hanbury-Brown and Twiss (HBT) setup containing a 50:50 multi-mode beamsplitter. For photon cross-correlation measurements the spectrally distinguishable exciton and biexciton emission (V-polarization) is spatially separated using two monochromators. In both cases (auto- and cross-correlation), coincidence measurements are performed using two fiber-coupled Silicon-based single-photon counting modules (SPCM) with an overall timing resolution of 350 ps in combination with time-correlated single-photon counting electronics with 4 ps time-bin width. In order to determine the efficiency of our twin-photon source from the detected count rates at the SPCMs, we measured the collection efficiency ε of our experimental setup to be $(0.95 \pm 0.05)$% following ref. 25. The indistinguishability of photons from the emitted biexciton-exciton pair is studied by means of Hong-Ou-Mandel (HOM)-type two-photon interference measurements via a Mach-Zehnder interferometer based on polarization maintaining fibers. A $\lambda/2$-waveplate allows to switch the polarization of photons in one of the interferometer arms, either being co- or cross-polarized with respect to photons in the other arm. The Mach-Zehnder interferometer in this work - in contrast to e.g. ref. 25 - was chosen symmetric with respect to the arm length, taking account for a negligible temporal delay between photon-twins.

**Twin-photon generation efficiency.** In order to quantify the efficiency of our twin-photon source, we introduced the parameter $\alpha = g^{(2)}_{auto}(0)/g^{(2)}_{cross}(0)$, describing the fraction of two-photon correlations due to twin-photon emission. This can be explained by considering the actual observables of our measurement in Fig. 2. In case of the cross-correlation, each start- and stop-trigger of the coincidence measurement can be attributed to the distinct detection of one XX and one X photon. The corresponding observable of this cross-correlation measurement can be expressed (for $\tau \geq 0$) by $g^{(2)}_{cross} = g^{(2)}_{XX-X}$. In case of the auto-correlation measurement on the degenerate H-polarized decay path, XX and X photons can be detected at both detectors. Thus the distinct time-order is lost and the correlation $g^{(2)}_{X-XX}$ from above is superimposed by the time-inverted correlation $g^{(2)}_{X-XX}$. Additionally, also the "true" auto-correlations $g^{(2)}_{X-X}$ and $g^{(2)}_{XX-XX}$ of exciton and biexciton, respectively, have to be taken into account. The complete two-photon correlation $g^{(2)}_{auto}$ for H-polarization thus reads ($\tau \geq 0$):

$$g^{(2)}_{auto} = \alpha\, g^{(2)}_{XX-X} + \beta\, g^{(2)}_{X-XX} + \gamma\, g^{(2)}_{X-X} + \delta\, g^{(2)}_{XX-XX}. \quad (1)$$

At zero delay-time ($\tau = 0$) only the term $g^{(2)}_{XX-X}$ on the right hand side of equation (1) has a non-zero contribution ($g^{(2)}_{XX-X}(0) > 1$), while in all other cases the QD has to be refilled with either one or two electron-hole pairs prior stop-photon detection and hence $g^{(2)}_{X-XX}(0) = g^{(2)}_{X-X}(0) = g^{(2)}_{XX-XX}(0) \equiv 0$. These considerations result in the expression $g^{(2)}_{auto}(0) = \alpha \times g^{(2)}_{cross}(0)$ and $\alpha$ is expected to be ¼ in case of equally distributed probabilities for all possible correlations. It follows, that one observes a preferred twin-photon emission, i.e. enhanced probability for the correlation $g^{(2)}_{XX-X}(0)$, if $\alpha > \frac{1}{4}$ (while $\alpha + \beta + \gamma + \delta = 1$). Thus, by introducing the parameter $\alpha$ we are able to quantify the efficiency of the cascade, despite the excitation-dependent exciton occupation discussed in previous reports[30,31].

**Twin-photon emission rate from CW experiments.** To deduce the twin-photon emission rate (TPR) collected via the microscope objective from the experimentally determined $\alpha$, one must take into account the measured count rates $n_{SPCM} = 103$ kHz at the SPCMs, the setup efficiency $\varepsilon = (0.95 \pm 0.05)\%$ for photons emitted into the first lens, and the photon extraction efficiency $\eta = (9 \pm 1)\%$ of the microlens (both, $\varepsilon$ and $\eta$ were measured independently according to ref. 25). At this point, we assume that the detected rate of photon twins at the SPCMs is negligible, such that the photon stream consists of single photons. These parameters at hand, we can calculate the photon rate emitted by the QD. According to our measurement $\alpha = 39\%$ of the two-photon correlations originate from the emission of photons twins (XX-X). The two-photon coincidences resulting from the remaining contributions (weighted by $\beta + \gamma + \delta$), however, result from photons in different excitation cycles (X-X, XX-XX, X-XX) and have to be counted as single photons. To be consistent and to calculate back to the contribution of single photon events one has to count them twice. This results in a probability of 24% for detecting a photon twin and 76% for the detection of single photons. With these values, we can calculate the twin-photon rate collected via the microscope objective:

$$TPR = \frac{n_{SPCM}}{\varepsilon \times \eta} \times 0.24 \times \eta^2 = (234 \pm 4) \text{ kHz},$$

where the quadratic dependence on $\eta$ was taken into account for photon twins.

**Two-photon interference visibility.** To extract the two-photon interference visibility from measured $g^{(2)}_{HOM}(\tau)$ traces, we first determined the peak area ratio $g^{(2)}_{HOM}(0) = A_0/A$, where $A_0$ corresponds to the area of the zero-delay peak and $A$ is the mean area of the peaks at $\tau \neq 0$. Even in the case of perfect indistinguishability between X and XX photons of the photon twins, one expects a finite contrast between the measurements in co- and cross-polarized configuration according to $g^{(2)}_{HOM,\parallel}(0) = \frac{1}{2} g^{(2)}_{HOM,\perp}(0)$. The respective coincidences in co-polarized configuration arise from the fact, that in 50% of all cases both photons of the exciton-biexciton pair will take the same path within the interferometer. Hence, they enter the second beam splitter at the same entrance port and thus cannot lead to two-photon interference. Consequently, the two-photon interference visibility has to be renormalized by a factor of 2 compared to the standard formula[39] according to $V = \frac{g^{(2)}_{\perp}(0) - g^{(2)}_{\parallel}(0)}{g^{(2)}_{\perp}(0) \times 1/2}$.

**Photon-number-resolving measurements.** For the photon-number-resolving (PNR) experiments, we employed a detection system based on a fiber-coupled transition-edge sensor (TES) operated in a cryogenic environment. The TES thereby acts as a highly sensitive calorimeter, which is able to detect smallest amounts of energy dissipated during photon absorption[40]. The detector is voltage-biased to heat up the electron system within its superconducting-to-normal-conducting transition (~152 mK) in the self-calibrated region. The implemented circuit allows for detecting a temperature increase, which causes a change in the resistance, ultimately leading to a detectable change in current. The latter is measured via an inductively coupled two-stage dc-superconducting quantum interference device (SQUID)[41]. For optimized absorption in the near infrared, the TES is embedded within a dielectric cavity[42], resulting in a detection efficiency of ~84% in the spectral region of interest for the detector used here. The TES/SQUID detector unit is mounted onto the cold stage of an adiabatic demagnetization refrigerator (ADR) stabilized at 100 mK.

For the PNR experiment, the emission of our twin-photon source is triggered by a pulsed diode laser (pulse duration ~80 ps) at a repetition rate of 1 MHz ($\lambda$ = 661 nm). The lower repetition rate is required in case of the PNR experiments, due to the relatively long thermal recovery time (~1 µs) after photon detection. The emission of the H-polarized decay path is spectrally filtered (bandwidth: 120 µeV) and coupled to the TES, using a single-mode fiber (Thorlabs 780HP) positioned right above the detector chip. In order to reduce contributions of background counts as far as possible, we first triggered our experiment with the detection of photons falling within a 220 ns wide time-window in succession of the laser trigger (taking into account the signal propagation time). This trigger mode enabled us to reduce the background counts down to ~3.6 twins/h and ~36 singles/h, caused by spurious detection events of ambient light photons entering the ADR via the optical fiber. Within a measurement period of 4.5 hours, we detect a total of 215 photon twins emitted by our QD source and we extract a twin-to-single photon ratio of "2/1" = $(1.81 \pm 0.05) \times 10^{-4}$ from the recorded histogram shown in Fig. 4c. Additionally, we determined the vacuum contribution by a second measurement, at which the laser sync output was used as a trigger. We determine a ratio of single-photon detection events to vacuum contribution of "1/0" = $1.1 \times 10^{-4}$ within an acquisition time of 18 min. In order to deduce the photon number distribution emitted by the QD from the detected ratios "2/1" and "1/0", we take a binomial distribution into account[43], where the number of independent Bernoulli trials is given by the photon number n = 0, 1, 2 according to the detection of zero, one and two photons, and the success probability of each Bernoulli trial is the product $\varepsilon_{PNR} \times \eta = 0.0504\%$ of the setup transmission $\varepsilon_{PNR} = (0.56 \pm 0.04)\%$ and the photon extraction efficiency of our microlens $\eta = (9 \pm 1)\%$. From the extracted probability for twin-photon emission $p_{twin} = 8.0\%$ we are able to calculate the triggered twin-photon emission rate collected via the microscope objective by $TPR = f \times p_{twin} \times \eta^2 = 80$ MHz $\times 0.080 \times 0.09^2 = (52^{+21}_{-12})$ kHz, by assuming an excitation rate of $f$ = 80 MHz. The complete procedure described above (PNR experiments and data analysis) was additionally carried out for a QD single-photon source, for a better comparison. The reconstructed photon number distributions resulting from these experiments are illustrated in Fig. 4d.

# Supplemenatary Information

## 1 Polarization-resolved measurements

For a quantitative analysis of the quantum dot's (QD's) bright exciton fine-structure splitting $\Delta E_{\text{FSS}}$, we applied fits to the spectra shown in Fig. 1d of the main article. We used four Lorentzian profiles corresponding to the linearly polarized emission lines $X_H$ and $X_V$ as well as $XX_H$ and $XX_V$ of the exciton- and biexciton-state, respectively. For carrying out the fits we made the following reasonable assumptions: Firstly, both excitonic and both biexcitonic components are assumed to have the same spectral linewidth $\gamma$: $\gamma_{X,H} = \gamma_{X,V}$ and $\gamma_{XX,H} = \gamma_{XX,V}$. Secondly, the energetic splitting of the excitonic components equals the splitting of the biexcitonic components: $\Delta E_X^{\text{H-V}} = \Delta E_{XX}^{\text{H-V}} = \Delta E_{\text{FSS}}$. Additionally, the ratio of the integrated intensities of exciton and biexciton is the same for both polarizations: $I_{X,H}/I_{XX,H} = I_{X,V}/I_{XX,V}$. The resulting relative spectral positions of the emission lines extracted from the fits to a total of 36 spectra are depicted in Fig. S1 in a histogram (upper panel) together with the original polarization- and energy-resolved contour plot (lower panel). The histograms reveal mean relative spectral positions of $\Delta E_{X,H} = (1.9 \pm 3.7)$ µeV and $\Delta E_{XX,H} = (-1.9 \pm 2.1)$ µeV for the H-polarized exciton and biexciton component, respectively, which coincide within their standard deviation. Hence, the photon-twins emitted in the H-polarized decay channel can be considered degenerate in energy and polarization. Further, taking into account the relative spectral positions of $\Delta E_{X,V}$ and $\Delta E_{XX,V}$ for the V-polarized decay channel, we can extract a fine-structure splitting of $\Delta E_{\text{FSS}} = (51 \pm 6)$ µeV.

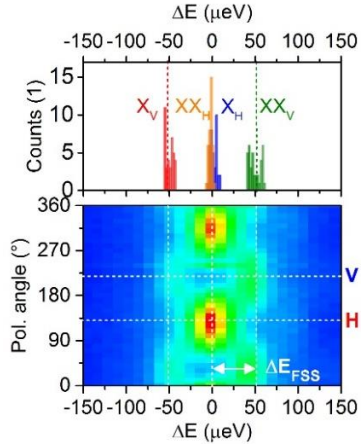

**Fig. S1: Quantitative analysis of the polarization resolved spectra.** The histogram (upper panel) shows the relative spectral positions of the four H- and V-polarized exciton (X) and biexciton (X) components extracted from a total of 36 spectra displayed in the lower panel.

## 2 Rate equation model

For a theoretical description of our twin-photon source, we model the QD as a four-level system constituted of a ground-state $|G\rangle$, two bright exciton states $|H\rangle$ and $|V\rangle$, and a biexciton state $|B\rangle$. The dynamics of the excitonic levels can thus be expressed by the following rate equations for the state-occupation probabilities $\rho$:

$$\frac{d}{dt}\rho_{BB} = -2\Gamma_B\rho_{BB} + P_B(\rho_{VV} + \rho_{HH})$$

$$\frac{d}{dt}\rho_{HH} = -\Gamma_X\rho_{HH} + \Gamma_B\rho_{BB} - P_B\rho_{BB} + P_X\rho_{GG}$$

$$\frac{d}{dt}\rho_{VV} = -\Gamma_X\rho_{VV} + \Gamma_B\rho_{BB} - P_B\rho_{BB} + P_X\rho_{GG}$$

$$\frac{d}{dt}\rho_{GG} = -2P_X\rho_{GG} + \Gamma_X(\rho_{HH} + \rho_{VV}),$$

where $\Gamma_B$ ($P_B$) and $\Gamma_X$ ($P_X$) correspond to the decay (pump) rates of the biexciton state and both exciton states. To connect the dynamics to the measurement, we assume within the far field approximation that the photon detection events are proportional to dipole excitations: $c_i^\dagger \equiv \sigma_{Bi} + \sigma_{iG}$, with $i = H,V$. Using this relation and the rate equations above, we can calculate the two-photon correlations $g^{(2)}(\tau \geq 0)$ in Formula (1) of the Methods section using the quantum regression theorem[1]:

$$g^{(2)}{}_{XX-X} = \frac{\langle \sigma_{BH}(0)\sigma_{HG}(\tau)\sigma_{GH}(\tau)\sigma_{HB}(0)\rangle}{\langle \sigma_{BH}\sigma_{HB}\rangle \langle \sigma_{HG}\sigma_{GH}\rangle}$$

$$= \frac{Na}{4\alpha P^3 \Gamma_B}\left(\beta_1 + \beta_2 e^{-(\Gamma_X + P)\tau} + \beta_3 e^{-(0.5\Gamma_X + \Gamma_B + 1.5P - 0.5\sqrt{\alpha})\tau} + \beta_4 e^{-(0.5\Gamma_X + \Gamma_B + 1.5P + 0.5\sqrt{\alpha})\tau}\right)$$

$$g^{(2)}{}_{X-XX} = \frac{\langle \sigma_{HG}(0)\sigma_{BH}(\tau)\sigma_{HB}(\tau)\sigma_{GH}(0)\rangle}{\langle \sigma_{BH}\sigma_{HB}\rangle \langle \sigma_{HG}\sigma_{GH}\rangle}$$

$$= \frac{e^{-(\Gamma_X + 2\Gamma_B + 3P + \sqrt{\alpha})\frac{\tau}{2}}}{2}\left(\frac{(1 - e^{\tau\sqrt{\alpha}})(\Gamma_X + 2\Gamma_B + 3P)}{\sqrt{\alpha}} - 1 - e^{\tau\sqrt{\alpha}}\right) + 1$$

$$g^{(2)}{}_{X-X} = \frac{\langle \sigma_{HG}(0)\sigma_{HG}(\tau)\sigma_{GH}(\tau)\sigma_{GH}(0)\rangle}{\langle \sigma_{HG}\sigma_{GH}\rangle^2}$$

$$= \frac{e^{-(\Gamma_X + 2\Gamma_B + 3P + \sqrt{\alpha})\frac{\tau}{2}}}{2\Gamma_B\sqrt{\alpha}}\left(-\Gamma_B\sqrt{\alpha}\left(1 + e^{\tau\sqrt{\alpha}}\right) - \left(1 - e^{\tau\sqrt{\alpha}}\right)\left((\Gamma_X - 2\Gamma_B)\Gamma_B + \Gamma_B P + P^2\right)\right) + 1$$

$$g^{(2)}{}_{XX-XX} = \frac{\langle \sigma_{BH}(0)\sigma_{BH}(\tau)\sigma_{HB}(\tau)\sigma_{HB}(0)\rangle}{\langle \sigma_{BH}\sigma_{HB}\rangle^2}$$

$$= \frac{e^{-(\Gamma_X + 2\Gamma_B + 3P + \sqrt{\alpha})\frac{\tau}{2}}}{2P\sqrt{\alpha}}\left(\left(1 - e^{\tau\sqrt{\alpha}}\right)(\Gamma_X + P)(-2\Gamma_B + P) - P\sqrt{\alpha}\left(1 + e^{\tau\sqrt{\alpha}}\right)\right) + 1$$

The following abbreviations were introduced for clarity:

$$P = P_X = P_B$$

$$N = \Gamma_X\Gamma_B + 2P\Gamma_B + P^2$$

$$\alpha = (\Gamma_X - 2\Gamma_B)^2 + 6\Gamma_X P - 4\Gamma_B P + P^2$$

$$\beta_1 = 4\Gamma_B P\alpha$$

$$\beta_2 = 2\alpha N$$

$$\beta_3 = \Gamma_X^3\Gamma_B - P^2(2\Gamma_B - P)\left(-2\Gamma_B + P + \sqrt{\alpha}\right) + \Gamma_X^2\left(-4\Gamma_B^2 + 6\Gamma_B + P^2 - \Gamma_B\sqrt{\alpha}\right)$$
$$\qquad + \Gamma_X\left(4\Gamma_B^3 + P^2\left(6P - \sqrt{\alpha}\right) + 2\Gamma_B^2\left(-2P + \sqrt{\alpha}\right) - 3\Gamma_B P\left(P + \sqrt{\alpha}\right)\right)$$

$$\beta_4 = \Gamma_X^3\Gamma_B + P^2(2\Gamma_B - P)\left(2\Gamma_B - P + \sqrt{\alpha}\right) + \Gamma_X^2\left(-4\Gamma_B^2 + 6\Gamma_B P + P^2 + \Gamma_B\sqrt{\alpha}\right)$$
$$\qquad + \Gamma_X\left(4\Gamma_B^3 + P^2\left(6P + \sqrt{\alpha}\right) + 2\Gamma_B^2\left(-2P + \sqrt{\alpha}\right) + 3\Gamma_B P\left(-P + \sqrt{\alpha}\right)\right)$$

$$Z = 4\alpha N$$

$$\langle \sigma_{BH}\sigma_{HB}\rangle = \frac{P^2}{N} = a$$

$$\langle \sigma_{HG}\sigma_{GH}\rangle = \frac{P\Gamma_B}{N}$$

In case of the photon cross-correlation on spectrally separable biexciton-exciton photons (see main article, Fig. 2a), the correlations $g_{X-XX}^{(2)}$ and $g_{XX-X}^{(2)}$ must be considered for negative and positive temporal delay τ, respectively. In case of the photon auto-correlation on spectrally degenerate biexciton-exciton photons (see main article, Fig. 2b), all four correlations from above get superimposed. To account for the timing resolution of the experimental setup, the theoretical correlation functions are convoluted with a Gaussian of 350 ps full-width at half maximum. Finally, the experimentally determined two-photon correlations in Fig. 2a and b as well as the inset of Fig. 3b are fitted using the derived model.

## 3 Indistinguishability of photon twins

For the two-photon interference visibility of photon twins observed in our experiment (cf. Fig. 4b) the following contributions can be considered: Firstly, pure dephasing in terms of an inhomogeneous spectral broadening of the QD emission results in a reduced wavepacket overlap[14]. A method to measure the corresponding time scale of spectral diffusion has been introduced very recently[44] and can be encountered in future by optimized QD growth[16] or electrically controlled microlens devices[35] to reduce electric field noise[17]. Secondly, in the case of interfering photons from the XX-X radiative cascade, exciton and biexciton state have different radiative lifetimes, independently measured to be $\tau_X = (1.77 \pm 0.05)$ ns and $\tau_{XX} = (0.95 \pm 0.08)$ ns, and the biexciton's radiative decay introduces an additional time jitter between the X and XX photons. This issue can be encountered in future by utilizing cavity effects to reduce the radiative lifetime via the Purcell-effect or to induce spontaneous two-photon emission[45]. Additionally, strictly resonant excitation can be applied in future, which eliminates any residual time jitter due to charge carrier relaxation.